\documentclass[10pt,twocolumn,letterpaper]{article}

\usepackage[pagenumbers]{wacv} 

\usepackage{graphicx}
\usepackage{amsmath}
\usepackage{amssymb}
\usepackage{booktabs}

\usepackage[table]{xcolor}
\usepackage{booktabs} 
\usepackage{multirow}
\usepackage{pifont} 
\usepackage[acronym]{glossaries} 
\usepackage{subcaption}
\usepackage{soul}

\usepackage{algorithm}
\usepackage[noend]{algpseudocode}

\algrenewcommand\algorithmicfunction{}
\algnewcommand{\LeftComment}[1]{\Statex \(\triangleright\) \textit{#1}}
\algnewcommand{\RightComment}[1]{\(\hfill //\) #1}
\algnewcommand{\BoldComment}[1]{\(\hfill \triangleright\) \underline{\emph{#1}}}
\algnewcommand{\CenterComment}[1]{\texttt{//#1}}

\definecolor{darkgreen}{RGB}{0, 153, 68}
\definecolor{darkred}{RGB}{165, 42, 42}
\definecolor{deepyellow}{RGB}{204, 204, 0}

\definecolor{Blue}{HTML}{0072B2}
\definecolor{Orange}{HTML}{E69F00}
\definecolor{SkyBlue}{HTML}{56B4E9}
\definecolor{BluishGreen}{HTML}{009E73}
\definecolor{Vermilion}{HTML}{D55E00}

\newcommand{\greencheck}{{\color{darkgreen}\ding{51}}}
\newcommand{\redcross}{{\color{darkred}\ding{55}}} 

\usepackage[pagebackref,breaklinks,colorlinks]{hyperref}

\usepackage[capitalize]{cleveref}
\crefname{section}{Sec.}{Secs.}
\Crefname{section}{Section}{Sections}
\Crefname{table}{Table}{Tables}
\crefname{table}{Tab.}{Tabs.}

\def\wacvPaperID{****} 
\def\confName{****}
\def\confYear{****}

\begin{document}

\title{Feasibility of Federated Learning from Client Databases with Different Brain Diseases and MRI Modalities}

\author{Felix Wagner$^1$ \and Wentian Xu$^1$ \and Pramit Saha$^1$ \and Ziyun Liang$^1$ \and Daniel Whitehouse$^2$ \and David Menon$^2$ \and
Virginia Newcombe$^2$ \and Natalie Voets$^3$ \and J. Alison Noble$^1$ \and Konstantinos Kamnitsas$^{1,4,5}$ \\
\small
$^1$Department of Engineering Science, University of Oxford;
\small
$^2$Department of Medicine, University of Cambridge;\\
\small
$^3$Nuffield Department of Clinical Neurosciences, University of Oxford;
\small
$^4$Department of Computing, Imperial College London;\\
\small
$^5$School of Computer Science, University of Birmingham\\
{\tt\small felix.wagner@eng.ox.ac.uk}
}
\maketitle
\renewcommand{\thefootnote}{}
\footnotetext{\vspace{-1.4em}Published as a conference paper at WACV 2025.}
\renewcommand{\thefootnote}{\arabic{footnote}}

\begin{abstract}
Segmentation models for brain lesions in MRI are typically developed for a specific disease and trained on data with a predefined set of MRI modalities. Such models cannot segment the disease using data with a different set of MRI modalities, nor can they segment other types of diseases. Moreover, this training paradigm prevents a model from using the advantages of learning from heterogeneous databases that may contain scans and segmentation labels for different brain pathologies and diverse sets of MRI modalities. Additionally, the confidentiality of patient data often prevents central data aggregation, necessitating a decentralized approach. 
Is it feasible to use Federated Learning (FL) to train a single model on client databases that contain scans and labels of different brain pathologies and diverse sets of MRI modalities? We demonstrate promising results by combining appropriate, simple, and practical modifications to the model and training strategy: Designing a model with input channels that cover the whole set of modalities available across clients, training with random modality drop, and exploring the effects of feature normalization methods.
\
Evaluation on 7 brain MRI databases with 5 different diseases shows that this FL framework can train a single model achieving very promising results in segmenting all disease types seen during training. Importantly, it can segment these diseases in new databases that contain sets of modalities different from those in training clients.
\
These results demonstrate, for the first time, the feasibility and effectiveness of using FL to train a single 3D segmentation model on decentralised data with diverse brain diseases and MRI modalities, a necessary step towards leveraging heterogeneous real-world databases. Code: \url{https://github.com/FelixWag/FedUniBrain}
\end{abstract}

\begin{figure}[t]
    \centering
    \includegraphics[width=0.95\columnwidth]{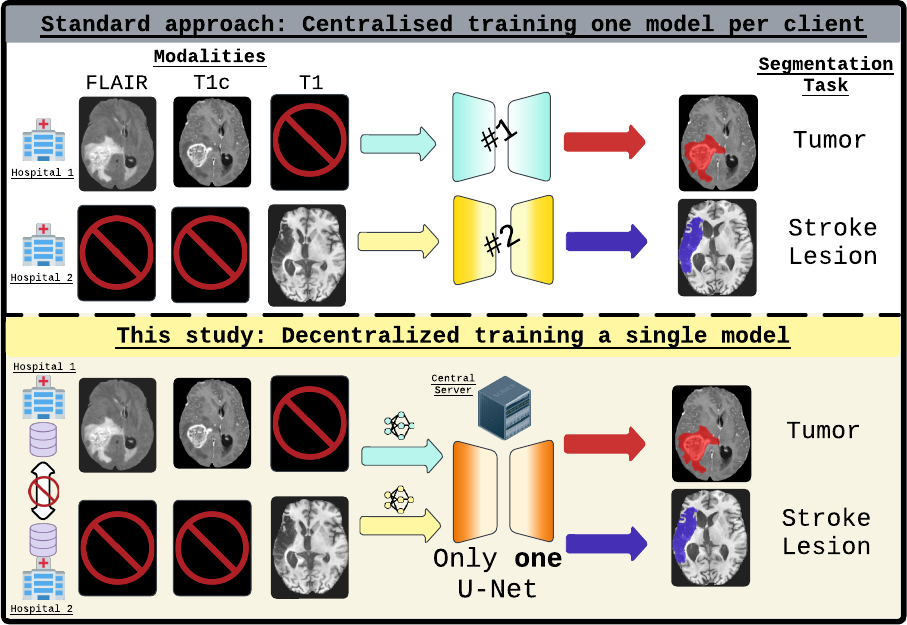}
    \caption{\textbf{FedUniBrain} (yellow box) enables the collaborative training of a single brain MRI segmentation model capable of segmenting multiple brain pathologies across multiple institutions without data sharing. \textbf{FedUniBrain} can train a model on data from institutions with varying sets of input MRI modalities and different brain diseases. In contrast, standard brain MRI segmentation models (grey box), can only process MRI scans with a specific set of MRI input modalities and are disease-specific.}
    \label{fig:ProblemOverview}
\end{figure}

\section{Introduction}

Federated Learning (FL) \cite{McMahan2017fedavg} is a promising approach for training machine learning models in healthcare, where patient privacy and data-sharing restrictions are critical. In medical imaging, multi-institutional collaborations are essential for training models that perform well across diverse patient populations \cite{shilo2020axes, peiffer2020machine}. However, privacy regulations prevent institutions from sharing confidential patient data, preventing the creation of an appropriately diverse centralised database. FL enables collaboratively model training across institutions without data sharing, allowing the use of larger, decentralized data. However, FL faces the challenges of data heterogeneity from different institutions, including variations in scanners, and disease specializations, complicating the decentralized optimization \cite{li2021fedbn, karimireddy2020scaffold}.
This study focuses on \textit{multi-modal} MRI segmentation, introducing additional heterogeneity in input modalities. Each MRI modality (e.g. FLAIR, T1, T2) provides complementary information about tissues and anatomies, which is important for diagnosing different diseases. For example, FLAIR is useful for detecting periventricular lesions in Multiple Sclerosis, T1c highlights enhancing tumour, while SWI highlights bleeding. Depending on equipment, disease specialization, and clinical protocols, institutions collect scans with different input modalities and brain pathologies. Therefore, training a unified MRI segmentation neural network across institutions with different sets of MRI modalities and diseases via FL introduces the challenge of handling varying sets of input modalities per institution, while being effective in segmenting multiple diseases. An additional challenge is enabling new clients joining the federation with previously unseen modality combinations and pathologies.
Traditionally, multi-modal MRI models for brain lesion segmentation are trained on specific disease data, acquired with a predefined protocol and set of MRI modalities \cite{kamnitsas2017efficient,kayalibay2017cnn,zhou2020one}. These models cannot segment other types of diseases, nor can they process scans with a different set of MRI modalities compared to the specific set in the training data.
It is an open question whether it is possible to effectively train a model with FL using decentralised databases, each with different types of pathologies and sets of MRI modalities.

To address these challenges, this study presents, \textbf{FedUniBrain} (\textbf{Fed}erated \textbf{Uni}versal \textbf{Brain} MRI Segmentation) a FL framework
for training a \textit{single} 3D segmentation model on diverse brain lesion databases, each with a different disease and set of MRI modalities (Fig.~\ref{fig:ProblemOverview}). This challenging setting involves three types of heterogeneities within the MRI data: \textbf{difference in scans}, \textbf{different segmentation tasks (brain pathologies)}, and \textbf{diverse sets of MRI input modalities} across clients. While many FL works have addressed heterogeneous medical imaging data, this is the first one to address all three types of heterogeneities combined, whereas previous works mainly focused on one type \textbf{and} assumed the same set of modalities per client, making it unclear how to even train a single model with varying input modalities per clients. Central questions for this study are: Is federated training of such a model feasible? Will it be beneficial compared to training disease-specific models? Will it result in a model that can segment new data with a different combination of modalities than the training data? Can a new client join the training with a modality and/or pathology not seen during training?

\noindent \textbf{Contribution:} This study develops a simple but practical solution for training via FL a single model that can process different sets of modalities and segment brain lesions of different types.
We show that this can be achieved by using a standard U-net, enhanced such that input channels can receive the whole set of unique modalities available across clients, and using random drop of modalities during training to prevent associations of modalities with specific pathologies. Additionally, due to the high data distribution shift across diverse databases, we present an analysis of the effects of feature normalization techniques.
Experimentation on 7 brain MRI databases with 5 different pathologies and 7 MRI modalities provides the following findings:

\begin{itemize}
    \item \underline{\textbf{Feasibility:}} This study is the first to show feasibility of Federated training of a \textbf{single} segmentation model across decentralised databases, with varying sets of input modalities and segmentation tasks
    \item \underline{\textbf{Knowledge Transfer Across Pathologies:}} This paper demonstrates effectiveness of FL for jointly learning to segment multiple pathologies, achieving performance that compares favorably to disease-specific models.
    \item \underline{\textbf{Segmenting new combination of modalities:}} This study shows that models trained with this paradigm can segment data from new databases with combination of modalities not seen during training.
    \item \ul{\textbf{Client with unseen modality and/or pathology joins the federation:}} We show that our framework is flexible and can handle clients that join the federation with a previously unseen modality and/or pathology.
\end{itemize}

\section{Related Work}
\noindent \ul{\textbf{Federated Learning on Heterogeneous Databases.}}
Many previous studies have explored and demonstrated the potential of FL for heterogeneous medical imaging data \cite{wagner2023staralign,li2019privacy,sheller2019multi,pati2022federated,hetFL,dinsdale2022fedharmony,yan2023label,jiang2022harmofl,zhang2022splitavg,hernandez2024review}. However, these works typically address heterogeneity with respect to scanner differences or label distribution, commonly using one imaging modality or the same set of modalities at each client. Previous works on FL specifically for brain MRI segmentation \cite{li2019privacy,sheller2019multi,pati2022federated} focused on \textbf{single-disease} scenarios, where all clients have access to the \textbf{same set of MRI input modalities}. These studies demonstrated that FL can effectively train a multi-modal brain MRI segmentation model without centralizing data. However, none of these methods are applicable when clients have diverse sets of input modalities. 
Only recently, a study \cite{xu2024feasibility} demonstrated for the first time the feasibility of training a segmentation neural network across multiple brain MRI databases with different diseases and sets of modalities, but in a centralised settings, it used a centralised database as a proof-of-concept. It did not address the practical challenge of medical data sharing restriction. In contrast, our study focuses on the challenging setting of decentralized FL, which has not been explored yet.

\noindent \ul{\textbf{Federated Multi-Task Learning.}} Centralised Multi-Task learning trains a model on multiple tasks to extract shared knowledge and generalize across tasks \cite{caruana1997multitask,crawshaw2020multi,ruder2017overview}. Recent Federated Multi-Task Learning \cite{cai2023many,lu2024fedhca2,chen2023fedbone} focuses on both data and \textit{task heterogeneity} in a federated setting (e.g. each client specializes on a different segmentation task), but none of these works is applied to medical imaging or applicable to multi-MRI modalities inputs.
In medical imaging, few papers explore Multi-Task FL. \cite{shen2021multi} studies FL for learning from data with different segmentation labels in CT. \cite{huang2022federated} tackles the problem of multi-task classification of mental disorders from MRI scans, but both of theses works only use one imaging modality as input. In summary, while FL with multi-modal MRI and FL for heterogeneous tasks have been studied; the combination of both, especially with different sets of modalities per client, has not been explored.

\noindent \ul{\textbf{Missing Modalities.}}
Learning from databases with varying sets of MRI modalities relates to research on missing modalities during inference \cite{havaei2016hemis,sharma2019missing,hu2020knowledge,zhou2021latent,wang2021acn}, often caused by clinical factors like acquisition protocols, costs or scanner availability. 
Most of these methods focus on disease-specific models with the same modalities used during training and in a centralized setting. FL with missing modalities is a recent research area but mainly focusing on text and image modalities \cite{saha2024examining,bao2023multimodal}, non-imaging data \cite{chen2022fedmsplit}, or require unimodal or complete multimodal data \cite{yu2023multimodal}. Notably, none of these approaches is applicable to varying MRI modalities to segment multiple tasks.
While addressing missing modalities is an important capability of our method, it is not the primary topic of this research.

\noindent \ul{\textbf{Comparison to Foundation Models.}}
Our study shares similarities with foundation models like SAM \cite{kirillov2023segment} and MedSAM \cite{ma2024segment} for medical imaging, which perform and generalize well to a variety of tasks. However, these models are not designed for 3D input data or multi-modality MRI input. Additionally, unlike our method, these models are interactive and require user input prompts. Therefore, this research is not directly applicable to our problem.

\begin{figure*}[t]
    \centering
    \includegraphics[width=0.75\textwidth]{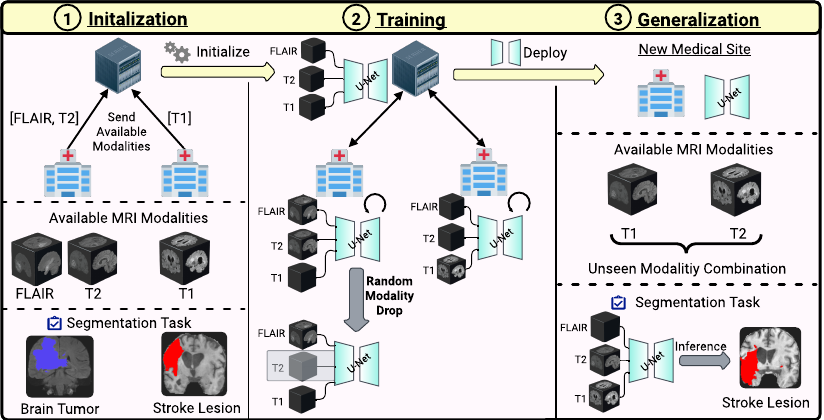}
    \caption{\textbf{Method overview.} (1) Initialization of model with distinct set of modalities across clients. (2) Federated training of the model with random modality drop. (3) Generalization to unseen clients with different input MRI modality combination.}
    \label{fig:Overview}
\end{figure*}

\section{Method}
\subsection{Problem setting: FL with Multi-Modal MRI and Multiple Tasks}\label{sec:ProblemSetting}
We consider \(C\) clients, where each client \(c\) 
(\(1\! \leq\! c\! \leq\! C\)) 
has a multi-modal MRI dataset \( \mathcal{D}_c = \{(\mathcal{X}_{ci}, y_{ci}) \}_{i=1}^{|\mathcal{D}_c|} \) with the set of MRI-modalities \(\mathcal{M}_c\), where \( \mathcal{X}_{ci} = \{ {x}_{ci}^{m} \}_{m=1}^{|\mathcal{M}_c|}\) denotes the multi-modal MRI image and \(y_{ci}\) the corresponding segmentation mask for sample \(i\). Here, each \(x_{ci}^{m} \in \mathbb{R}^{w \times h \times d}\), represents a 3D image, where \(w, h\) and \(d\) denote the width, height, and depth respectively. Furthermore, \( \mathcal{M} = \bigcup_{i=1}^{C} M_i \) denotes the set that represents all unique modalities across all clients and \(|\mathcal{M}|\) is the total number of unique modalities across all datasets. Each client solves a specific segmentation task \(\mathcal{T}_c\).
Our goal is to train, without data sharing, a \textbf{single} global model parameterized by \( \theta \), that performs well on all tasks and clients. The global objective is to minimize the average of the client-local loss functions:
\begin{equation}
\min_{\theta} [\mathcal{L}(\theta) := \frac{1}{C} \sum_{c=1}^{C} \mathcal{L}_c(\theta; \mathcal{D}_c; \mathcal{T}_c)],
\end{equation}
where \(\mathcal{L}_c\) denotes the local loss function of client \(c\).
As loss functions, we use a weighted combination of Dice and pixel-wise Binary Cross-Entropy (BCE) losses:
\begin{equation}
    \mathcal{L}_c = \alpha \mathcal{L}_{Dice} + (1-\alpha) \mathcal{L}_{BCE}
\end{equation}
In the standard FL setting with \(E\) communication rounds, a global server initializes a global model and communicates it to every client. In each round, each client performs \( \tau \) iterations of updates on its local data and loss function \(\mathcal{L}_c(\theta; \mathcal{D}_c; \mathcal{T}_c)\). Afterwards, every client sends their updated model back to the server. The server combines these updates, typically by averaging (FedAvg), into a new global model before starting a new communication round \cite{McMahan2017fedavg}. This allows decentralised training without data sharing. 

\noindent \textbf{Difference to other FL settings:} We highlight that, in comparison to the standard FL setting, the setting studied herein varies not only with respect to input space with \(|\mathcal{M}|\! >\! 1\), considering multiple modalities, but also the label space \(|\bigcup_{i=1}^{C} \mathcal{T}_i|\! >\! 1\), segmenting different brain pathologies per client. Moreover, this setting considers different sets of modalities per client, an aspect mostly overlooked in FL. Comparison of different FL settings is shown in Tab.~\ref{tab:FLComparison}.

\begin{table}[htbp]
\centering
\caption{Previous FL settings are not designed to simultaneously handle different tasks, multiple MRI input modalities, missing modalities, and varying sets of modality combinations across clients. FedUniBrain handles all these aspects simultaneously.}
\label{tab:FLComparison}
\resizebox{\columnwidth}{!}{
\begin{tabular}{lcccc}
\toprule
\textbf{Setting} & \textbf{Multi Task} & \textbf{Multi MRI Mod.} & \textbf{Missing Mod.} & \textbf{\textcolor{BluishGreen}{Varying} Mod. Comb.} \\
\midrule
Standard FL & \redcross & \redcross & \redcross & \redcross \\
Multi-Task FL & \greencheck & \redcross & \redcross & \redcross \\
Multi-MRI Mod. FL & \redcross & \greencheck & \redcross & \redcross \\
\textbf{FedUniBrain} & \greencheck & \greencheck & \greencheck & \greencheck \\
\bottomrule
\end{tabular}
}
\end{table}

\subsection{Unified Model for Multiple Segmentation Tasks}\label{sec:UnifiedModelTraining}

\noindent \textbf{Algorithm:} An overview of our simple yet effective algorithm is shown in Fig.~\ref{fig:Overview}. Initially, the central server receives a list of available MRI modalities from each client. Afterwards, the server initializes a model architecture that accepts the set of unique MRI modalities \(\mathcal{M}\) as input, and the FL algorithm starts (Sec.~\ref{sec:ProblemSetting}). During training, each client's input modalities are extended to the whole set of unique MRI modalities, with each missing modality filled with zeroes, and modality drop is applied. Supplementary Alg.~\ref{alg:FedUniBrain} contains the full algorithm for further details.

\noindent \textbf{Handling Different Sets of Modalities Across Clients:} Since every client comes with a different combination of MRI modalities, the first challenge arising is to define a single neural network architecture that can process different sets of input modalities. To address this, we augment each client's input images to cover the full set of $|\mathcal{M}|$ distinct MRI modalities across all clients. For each client \(c\), any missing modalities in the set \(\mathcal{M} - \mathcal{M}_c \) are filled with zeroes. Therefore, the input space for each client \(c\) is extended from \( \mathcal{X}_c \in \mathbb{R}^{|\mathcal{M}_c| \times w \times h \times d} \) (\textit{only client-specific modalities}) to \( \mathcal{X}_c \in \mathbb{R}^{|\mathcal{M}| \times w \times h \times d} \), where \(|\mathcal{M}|\) is the total number of modalities across all clients. Consequently, our model accepts the set of $|\mathcal{M}|$ distinct MRI modalities as input and can be represented as a function \( f(\cdot; \theta): \mathbb{R}^{|\mathcal{M}| \times w \times h \times d} \to \mathbb{R}^{w \times h \times d} \), capable of processing each client's input modality combination and outputting a 3D segmentation mask, ensuring compatibility across all clients.

\noindent \textbf{Modality Drop for Pathology Disassociation \& Missing Modalities:} To develop a model that generalises to modality combinations not seen during training, we employ \textbf{modality drop}, where we randomly set input modalities to zero during training. Therefore, we define a probabilistic masking function \(Z(\mathcal{X}, \phi)\) where each input modality \(x^m \in \mathcal{X}\) is set to zero with probability \(\phi\). We implement the following approach: During training, for each sample, we consider the set of initially available input modalities \(\mathcal{M}_c\) from the corresponding database. We then sample a random integer \( r \in [1, |\mathcal{M}|]\), ensuring that at least one modality remains. Next, we randomly select \(r\) number of modalities from \(\mathcal{M}_c\) to keep, setting the remaining modalities blank. This prevents the model from associating specific modality combinations with specific pathologies, ensuring it can accurately segment known pathologies in unseen clients with modality combinations not seen during training. Furthermore, it also enhances the model's robustness against missing modalities during testing, as shown in our evaluation.

\begin{figure*}[h]
  \centering
  \begin{subfigure}[t]{0.68\linewidth}
  \includegraphics[width=\linewidth]{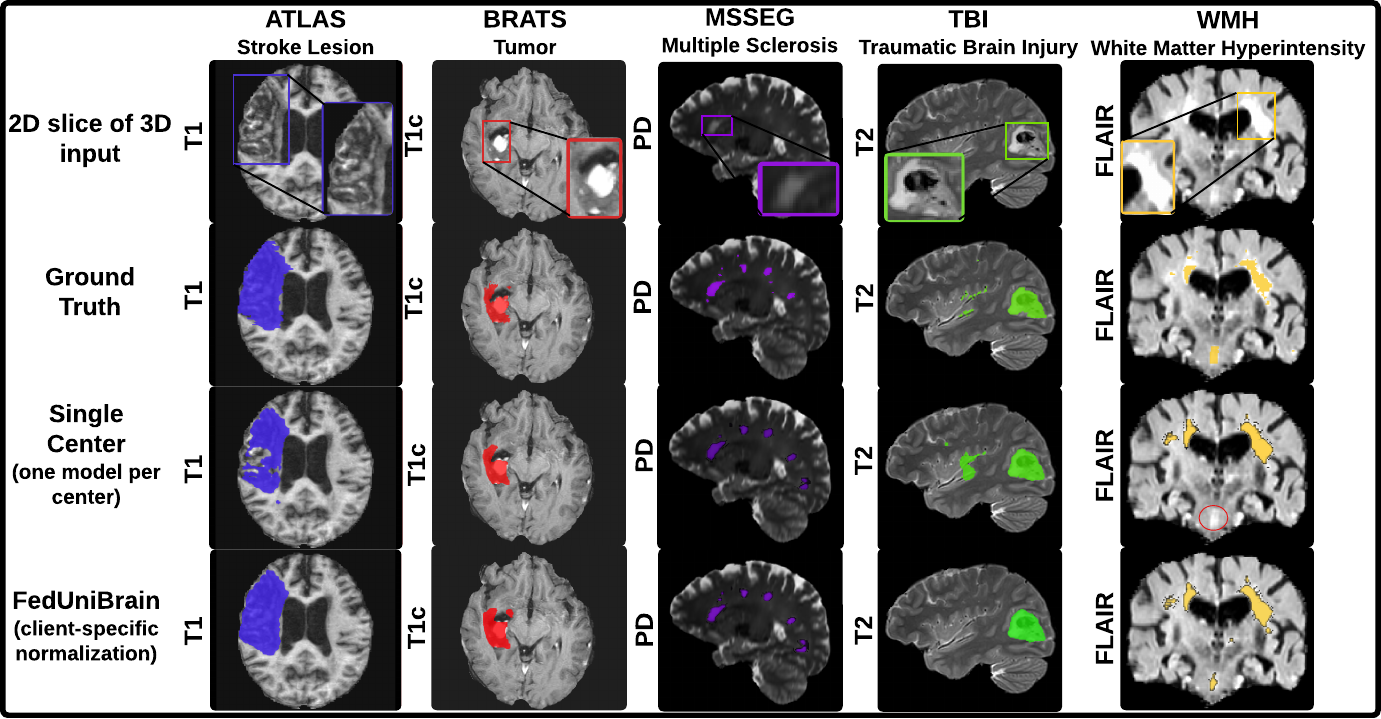}
    \caption{Segmentation results on source clients.}
    \label{fig:segmentations-sub-1}
  \end{subfigure}
  \hfill
  \begin{subfigure}[t]{0.3154\linewidth}
    \includegraphics[width=\linewidth]{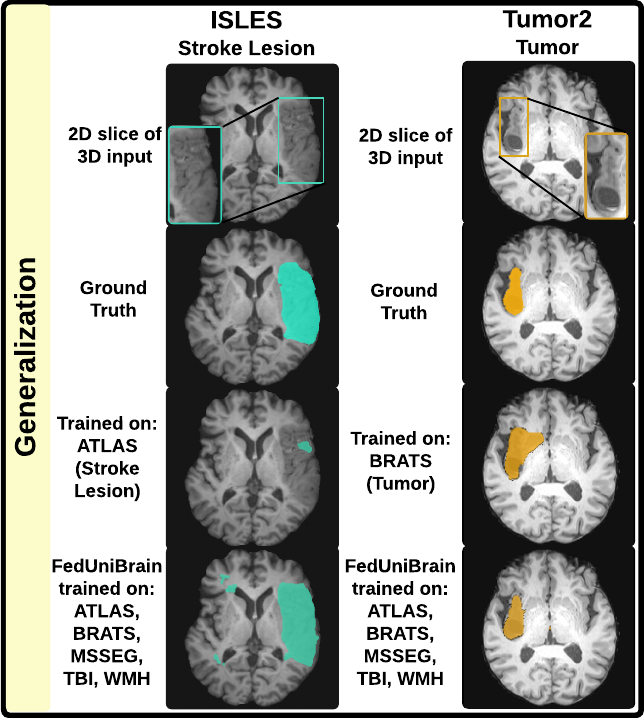}
    \caption{Generalization to unseen sets of modalities.}
    \label{fig:segmentations-sub-2}
  \end{subfigure}
  \caption{\textbf{Example segmentation results for databases with different pathologies.} (a) Segmentation outputs for five different databases, each having a different pathology to segment and different modality combination. Shown are 2D slices of the 3D input, the ground truth, predictions from individually trained U-Nets (on each database separately), and our FedUniBrain (trained on ATLAS, BRATS, MSSEG, TBI and WMH) model predictions. For each database we show a different modality. (b) Segmentation outputs from a FedUniBrain model trained on ATLAS, BRATS, MSSEG, TBI, and WMH, tested on excluded ISLES and Tumor2 databases with unseen modality combinations. Also, results from disease-specific models are shown: one trained on the BRATS (Tumor) and evaluated on Tumor2, and another trained on ATLAS and evaluated on ISLES for stroke lesions.}
  \label{fig:short}
\end{figure*}

\noindent \textbf{Feature Normalization for Heterogeneous Clients:} Batch Normalization (BN) \cite{ioffe2015batch} is an indispensable feature normalization technique that showed various benefits in training deep neural networks \cite{ioffe2015batch, santurkar2018does}. BN normalizes the features for each mini-batch as follows:
\begin{equation}
    \text{BN}(\mathbf{x}) = \gamma \frac{\mathbf{x} - \mu}{\sqrt{\sigma^2 + \epsilon}} + \beta,
\end{equation}
where \(\mu\) and \(\sigma\) represent the mean and variance computed over the mini-batch, while \( \gamma \) and \( \beta \) are learnable parameters. Term \(\epsilon\) is used for numerical stability. During inference \(\mu\) and \(\sigma\) are replaced with running averages \(\hat{\mu}\) and \(\hat{\sigma}\) computed during training to avoid reliance on mini-batches. Despite its success, it has been shown that BN does not perform well in multi-domain training as these statistics are not representative for individual domains \cite{LI2018109}. In our setting, each client's data comes from a different domain, with different pathologies, modalities and scanners, which increases the difficulty of training a model due to high data heterogeneity. This introduces a challenge when using BN for our setting.

To tackle this, we keep client-specific BN parameters and statistics (\( \gamma_c \), \( \beta_c \) and \(\hat{\mu}_c\), \(\hat{\sigma}_c\)) for each client \(c\), and exclude them from FL's model averaging step, as proposed in \cite{andreux2020siloed,li2021fedbn}. Our experiments show that this method improves performance on the training (\textbf{source}) databases. However, it is unclear how to apply the source-specific BN layer parameters/statistics to unseen (\textbf{target}) clients during inference. One way involves using averages of learned BN parameters, $\bar{\gamma}\!=\!\frac{1}{C}\sum_{c=1}^{C}{\gamma_c}$ and $\bar{\beta}\!=\!\frac{1}{C}\sum_{c=1}^{C}{\beta_c}$, and estimating target-specific running average statistics $\hat{\mu}_t$ and $\hat{\sigma}_t$ with a forward pass over the target database. The latter poses the disadvantage that a pre-collected target database is needed prior to inference. 
Therefore, in our experiments, we also explore alternative feature normalization methods for this training framework with inherently high data heterogeneity. 
Specifically, we investigate the recently proposed Normalization Free (NF) technique \cite{zhuang2024fedwon} that uses re-parameterized convolutional layers instead of explicitly normalizing statistics of feature activations, Group Norm (GN) \cite{wu2018group}, and Instance Norm (IN) \cite{ulyanov2016instance}. These approaches do not require target data to estimate the target statistics, alleviating this requirement of keeping client-specific BN layers. Empirical analysis (Sec.~\ref{sec:experiments}) also shows that this results in methods that generalize better to unseen clients. However, it slightly decreases performance on source clients. Therefore, the choice of feature normalization method in this framework should be chosen based on the target application for the trained model (generalization to new domains or personalization to source domains) and whether data are expected to have been collected a-priori from the target domain for estimating BN statistics.

\noindent \textbf{Client with unseen modality joins:} In a realistic FL scenario, a new client may join a federation at a later training stage, by contributing data to further improve a pre-trained model. If their data includes a new modality that was not in the originally available set $\mathcal{M}$, can we learn the new modality and data? To handle this scenario, we make the following model modification to process the new modality concatenated along with already known modalities as multi-channel input images. 
Assume a pre-trained network before the new client joins, and the first convolutional layer weights \(\mathbf{W}_1 \!\in\! \mathbb{R}^{o \times |\mathcal{M}| \times K_h \times K_w \times K_d} \), where \(o\) is the number of output filters, \(K_h, K_w, K_d\), the height, width, and depth of the kernel, respectively. 
To add the new modality, we extend dimensionality of \(\mathbf{W_1}\) to get \({\mathbf{W'_1} \in \mathbb{R}^{o \times (|\mathcal{M}|+1) \times K_h \times K_w \times K_d}}\). 
We then initialize the weights of the existing channels with the pre-trained weights: 
${\mathbf{W'_1}[:,i,:,:,:] = \mathbf{W_1}[:,i,:,:,:], \forall \, i \in [1, |\mathcal{M}|].}$
For the new modality, we randomly select one of the pre-existing modalities, sampling \(k \in [1, |\mathcal{M}|]\), and use its pre-trained weights to initialize the weights associated with the newly added modality:
${\mathbf{W'_1}[:,|\mathcal{M}|+1,:,:,:] = \mathbf{W_1}[:,k,:,:,:] }$. All network parameters, including $\mathbf{W'_1}$, are then trained as normal with data from all clients.

\begin{table*}[ht]
\centering
\caption{Separate models are trained and evaluated on each database (\textbf{single center}), with only BN results shown as they yield the best results. In a federated setting, \textbf{FedUniBrain} is trained jointly on ATLAS, BRATS, MSSEG, TBI, and WMH with various feature normalization methods. Results with/out modality drop are also reported. Results (Dice score \%) show the feasibility of federated training a unified model across multiple diverse brain databases, with performance comparable to or slightly outperforming disease-specific models.}
\label{tab:joint_training_results}
\resizebox{0.77\textwidth}{!}{
\begin{tabular}{@{}lccccccc|c@{}}
\toprule
\textbf{Method} & \textbf{Mod. Drop} & \textbf{Norm} & \textbf{ATLAS} & \textbf{BRATS} & \textbf{MSSEG} & \textbf{TBI} & \textbf{WMH} & \textbf{Average} \\ 
\midrule
Single Center & \redcross & BN & 52.8 & 91.9 & 68.3 & 56.1 & 71.5 & 68.1\\
FedUniBrain (client spec. BN params) & \redcross & BN & 50.9 & 91.8 & 70.2 & 55.8 & 74.8 & 68.7 \\
\midrule
Single Center & \greencheck & BN & 52.8 & 91.7 & 66.7 & 53.9 & 69.1 & 66.8 \\
FedUniBrain (avg. BN params) & \greencheck & BN & 43.1 & 89.4 & 63.2 & 51.4 & 67.3 & 62.9 \\
FedUniBrain & \greencheck & IN & 49.0 & 90.7 & 66.6 & 52.9 & 71.0 & 66.0 \\
FedUniBrain & \greencheck & GN & 48.1 & 91.0 & 67.3 & 53.7 & 70.4 & 66.1 \\
FedUniBrain & \greencheck & NF & 51.2& 91.1& 64.8 & 54.4 & 72.3 & 66.8 \\
FedUniBrain (client spec. BN params) & \greencheck & BN & 54.5 & 91.8 & 69.1 & 56.2 & 73.7 & \textbf{69.1} \\
\bottomrule
\end{tabular}
}
\end{table*}

\section{Experiments \& Results}
\label{sec:experiments}

\subsection{Datasets \& Prepocessing}
We use 7 annotated brain lesion MRI databases, each with different MRI modalities and segmentation tasks, collected from different sites: \textbf{ATLAS} v2.0 \cite{liew2022large} for stroke lesion, \textbf{BRATS} \cite{bakas2017advancing} via the Medical Segmentation Decathlon \cite{antonelli2022medical} for tumor, \textbf{MSSEG} \cite{commowick2018objective} for multiple sclerosis, \textbf{TBI} from our site for traumatic brain injury, \textbf{WMH} \cite{AECRSD_2022} for White Matter Hyperintensity, \textbf{ISLES} 2015's SISS subtask \cite{maier2017isles} for stroke lesion, and \textbf{Tumor2} for tumour segmentation from our institution. The dataset is divided into training/validation sets as follows: BRATS (444/40), MSSEG (37/16), ATLAS (459/195), TBI (156/125), WMH (42/18), ISLES (20/8), and Tumor2 (40/17). For the zero-shot generalization experiments (Sec.~\ref{sec:discussion}, Tab.~\ref{tab:generalization}), all samples are used to evaluate the model. For the Oracle results (Tab.~\ref{tab:generalization}), where the model is trained on the target database and not all samples can be used for evaluation, we apply 3-fold cross-validation, ensuring each sample is used for evaluation once. Tab.~\ref{tab:database_modalities} gives an overview of the available modalities. 
We resample the images to 1x1x1 mm, skull-strip, apply z-score intensity normalization, and combine all lesion types across databases into a single `lesion' label.

\begin{table}[h]
\centering
\caption{\textbf{MRI databases overview.} Shown are the sets of modalities in each database and the corresponding segmentation tasks.}
\label{tab:database_modalities}
\resizebox{\columnwidth}{!}{%
\begin{tabular}{lccccccc|l}
\toprule
\multirow{2}{*}{\textbf{Database}} & \multicolumn{7}{c|}{\textbf{Modalities}} & \multicolumn{1}{c}{\multirow{2}{*}{\textbf{Disease/Task}}} \\
\cmidrule{2-8}
& \textbf{T1} & \textbf{T1c} & \textbf{FLAIR} & \textbf{T2} & \textbf{PD} & \textbf{SWI} & \textbf{DWI}\\
\midrule
BRATS & \greencheck & \greencheck & \greencheck & \greencheck & & & & Tumor\\
MSSEG & \greencheck & \greencheck & \greencheck & \greencheck & \greencheck & & & Multiple Sclerosis \\
ATLAS & \greencheck &  &  & & & & & Stroke Lesion\\
TBI   & \greencheck & & \greencheck & \greencheck &  & \greencheck & & Traumatic Brain Injury\\
WMH   & \greencheck & & \greencheck & & & & & White Matter Hyperintensity\\
\midrule
ISLES & \greencheck & & \greencheck & \greencheck & & & \greencheck & Stroke Lesion\\
Tumor2 & \greencheck & & & & & & & Tumor\\
\bottomrule
\end{tabular}
} 
\end{table}

\subsection{Experimental Setup}\label{sub:Setup}
All experiments use the Residual U-Net~\cite{zhang2018ResUnet} as backbone, Adam optimizer, and \(\alpha \!=\! 0.8\) for our weighted loss function. We linearly scale the learning rate from 0 to \(10^{-3}\) over 50 communication rounds, keep it constant until communication round 150, and then linearly decay it. We set \(\tau\!=\!58\) (equivalent to 1 epoch for the largest database, ATLAS), \(E\!=\!300\) (or 300 epochs for centralised training), and batch size of 8. The Dice score on the validation set is reported after the last epoch. For FL, we study the effect of feature normalisation methods, assessing BN when all its parameters are aggregated as per the standard FedAvg algorithm \cite{McMahan2017fedavg}, and, separately, using FedBN \cite{li2021fedbn} to keep client-specific BN parameters (cf. Sec.~\ref{sec:UnifiedModelTraining}).
We also assess methods that do not rely on client-specific normalization layers: GN \cite{wu2018group}, IN \cite{ulyanov2016instance}, and NF \cite{zhuang2024fedwon}. To strengthen our findings, we tested the statistical significance of the main results for superiority or non-inferiority, as appropriate. A detailed analysis can be found in the Supplementary (Sec.~\ref{sup:sec:stats}).

\subsection{Results \& Discussion}\label{sec:discussion}

\noindent \ul{\textbf{Unified Model Improves Performance Across Databases}}

\noindent Experimental results in Tab.~\ref{tab:joint_training_results} prove that
it is feasible to use Federated Learning to train a single model jointly over client databases with diverse diseases and sets of modalities and reach performance levels of database-specific models. When learning client-specific BN parameters, FedUniBrain
matches or even surpasses single center training, with statistical superiority for ATLAS (p=0.03) and WMH (p$<$0.001) and non-inferiority for BRATS (p$<$0.001), MSSEG (p=0.011), and TBI (p$<$0.001), highlighting the benefits of knowledge transfer between brain diseases. Training with modality drop and client-specific BN layers leads to the highest overall results and performs better than using averaged BN parameters, demonstrating the benefits of client specific BN layers for source clients.
\
Visual examples of training a single model per client compared to FedUniBrain are shown in Fig.~\ref{fig:segmentations-sub-1}. Besides its practicality over disease-specific models, the unified model may have improved generalisation abilities on unseen databases, which are assessed in follow-up experiments.

\noindent \ul{\textbf{Modality Drop for Robustness to Missing Modalities}}

\noindent Our goal is to train a model that can segment pathologies even when modality combinations differ from those in the original training databases. This requires robustness against missing modalities. Results in Tab.~\ref{tab:modality_drop} show that training with modality drop is effective for this purpose. 
When training with modality drop, the Dice score decreases by only 3.9\% when evaluating with missing modalities compared to all modalities, versus 21.3\% decrease without modality drop.  
Therefore, the following experiments focus on models trained with modality drop.

\begin{table*}[h]
\centering
\caption{Assessing generalization to unseen databases, ISLES and Tumor2, each with modality combinations not seen during training (Dice score \% reported):
Separate models were trained on ISLES and Tumor2 and evaluated on the same database as Oracle baselines. We train task-specific baselines on stroke database ATLAS, then evaluate on ISLES, and train on tumor database BRATS, then evaluate on Tumor2, as generalization baselines. Centralised training across all 5 databases (MultiUNet) is also shown.
FedUniBrain is then trained with FL across 5 client databases (ATLAS, BRATS, MSSEG, TBI, WMH) using varying normalization methods, then evaluated on ISLES and Tumor2. 
The studied FL paradigm a) can reach performance similar to centralised training, even Oracles; b) generalizes to unseen databases better than task-specific models; c) IN and NF that don't learn client-specific parameters generalize best.}
\label{tab:generalization}
\resizebox{0.79\linewidth}{!}{
\begin{tabular}{@{}lccccccc@{}}
\toprule
\textbf{Method} & \textbf{Mod. Drop} & \textbf{Norm.} & \textbf{Needs Targ. Data} & \multicolumn{1}{c}{\textbf{ISLES}} & \multicolumn{1}{c}{\textbf{Tumor2}} & \textbf{Average} \\
\midrule
\multicolumn{1}{c}{\cellcolor{blue!10}} & \redcross & BN & Yes & 48.5 & 75.9 & 61.9 \\
\multicolumn{1}{c}{\cellcolor{blue!10}} & \redcross & IN & Yes & 53.2 & 76.5 & 64.9 \\
\multicolumn{1}{c}{\multirow{-3}{*}{\cellcolor{blue!10}\begin{tabular}[c]{@{}c@{}} Oracle:\\ Disease specific models trained on target only\end{tabular}}} & \redcross & NF & Yes & 54.8 & 73.4 & 64.1 \\
\midrule
Stroke-specific model (Trained on ATLAS (T1)) & \redcross & IN & No & 5.3 & - & \\
Tumor-specific model (Trained on BRATS) & \greencheck & IN & No & - & 69.1 & \multirow{-2}{*}{37.2} \\
\midrule
Centralised (MultiUNet \cite{xu2024feasibility}) & \redcross & IN & No & 50.7 & 58.1 & 54.4 \\
Centralised (MultiUnet \cite{xu2024feasibility}) & \greencheck & IN & No & 55.5 & 72.2 & 63.9 \\
\midrule
FedUniBrain (avg. BN params) & \redcross & BN & No & 53.8 & 0.0 & 26.9 \\
FedUniBrain (avg. BN params) & \greencheck & BN & No & 54.5 & 68.0 & 61.3 \\
FedUniBrain & \redcross & IN & No & 48.3 & 59.1 & 53.7 \\
FedUniBrain & \greencheck & IN & No & 55.3 & 72.7 & \textbf{64.0} \\
FedUniBrain & \redcross & NF & No & 54.6 & 31.1 & 42.9 \\
FedUniBrain & \greencheck & NF & No & 52.8 & 72.1 & 62.5 \\
FedUniBrain (client spec. BN params) & \redcross & BN & Yes & 50.2 & 36.8 & 43.5 \\
FedUniBrain (client spec. BN params) & \greencheck & BN & Yes & 49.9 & 70.1 & 60.0 \\
\bottomrule
\end{tabular}
}
\end{table*}

\begin{table}[ht]
\centering
\caption{Results from training FedUniBrain jointly on ATLAS, BRATS, MSSEG, TBI, and WMH using client-specific BN layers, with and without modality drop. Evaluation is done on the same databases with all modalities and by randomly dropping modalities during testing (missing modalities). 
Training with modality drop enables good performance with different sets of modalities at test time. Average Dice score over all databases is shown.}
\label{tab:modality_drop}
\resizebox{\columnwidth}{!}{ 
\begin{tabular}{@{}lcccc@{}}
\toprule
\textbf{Training Method} & \multicolumn{2}{c}{\textbf{Evaluation (Dice (\%))} $\uparrow$}                    & \multicolumn{1}{c}{ \(\Delta\) \textbf{Dice} $\downarrow$} \\
\cmidrule(lr){2-3} & All Modalities & Missing Modalities &  \\ 
\midrule
No Modality Drop & 68.7 & 47.4     & 21.3                   \\
Modality Drop & \textbf{69.1} & \textbf{65.2} & \textbf{03.9}                   \\ 
\bottomrule
\end{tabular}
} 
\end{table}

\noindent \ul{\textbf{Zero-Shot Generalization to Unseen Modality Combinations}}

\noindent We assess the generalization capabilities to unseen databases with modality combinations not seen during training by evaluating on ISLES and Tumor2. Related results are shown in Tab.~\ref{tab:generalization}. Remarkably, despite not having seen any data from ISLES or Tumor2 during training, the best federated model's average Dice score is only 0.9\% lower than the Oracle models, which were trained on the specific target databases.
FedUniBrain trained with FL and IN across 5 clients reaches performance levels of centralized training using the 5 databases (MultiUnet, with IN as it performed best), which is confirmed by our statistical non-inferiority tests on ISLES (p=0.003) and Tumor2 (p$<$0.001). This demonstrates feasibility of decentralised training with such diverse brain lesion MRI databases.

Traditionally, models are trained on databases specific to a disease and evaluated on unseen clients with the same disease, as we trained with ATLAS (for stroke lesion) and evaluated on ISLES (stroke database), and trained with BRATS (for brain tumors) and evaluated on Tumor2 (tumor database), setting this as baseline. Both are only evaluated using T1 since it is the only modality common between training and test databases. For the BRATS model, we trained with modality drop to allow inference on just T1 in the Tumor2 database. 
The remaining non-Oracle results for ISLES are evaluated on all ISLES modalities except DWI, as DWI is unique to ISLES and was not seen during training. 
Federated joint training achieves a performance gain of 47.5 on ISLES and 3.0 on Tumor2, reinforcing the benefits of joint training across multiple MRI databases compared to pathology specific models. 

Training with modality drop effectively disassociates specific modality combinations with certain pathologies. FedUniBrain with modality drop tends to show improved average performance over the two new databases with modality combinations not seen during training with visual results shown in Fig.~\ref{fig:segmentations-sub-2}. There is a large improvement for Tumor2 in all settings. Similarly, for ISLES, there is a general trend of improvement. Stochasticity in federated optimization may explain the few cases where modality drop made federated models perform slightly worse.

Non-client specific feature normalization, particularly IN, gives the best generalization results, with NF giving the second-best results. Therefore, these approaches are preferable for generalization compared to averaging client-specific BN layers, and estimating running statistics from target data, which yield the worst results. Additionally, the estimation requires unlabeled data the from unseen client.

\iftrue
\begin{figure*}[htbp]
    \centering
    \begin{subfigure}[t]{0.31\textwidth}
        \centering
        \includegraphics[width=\linewidth]{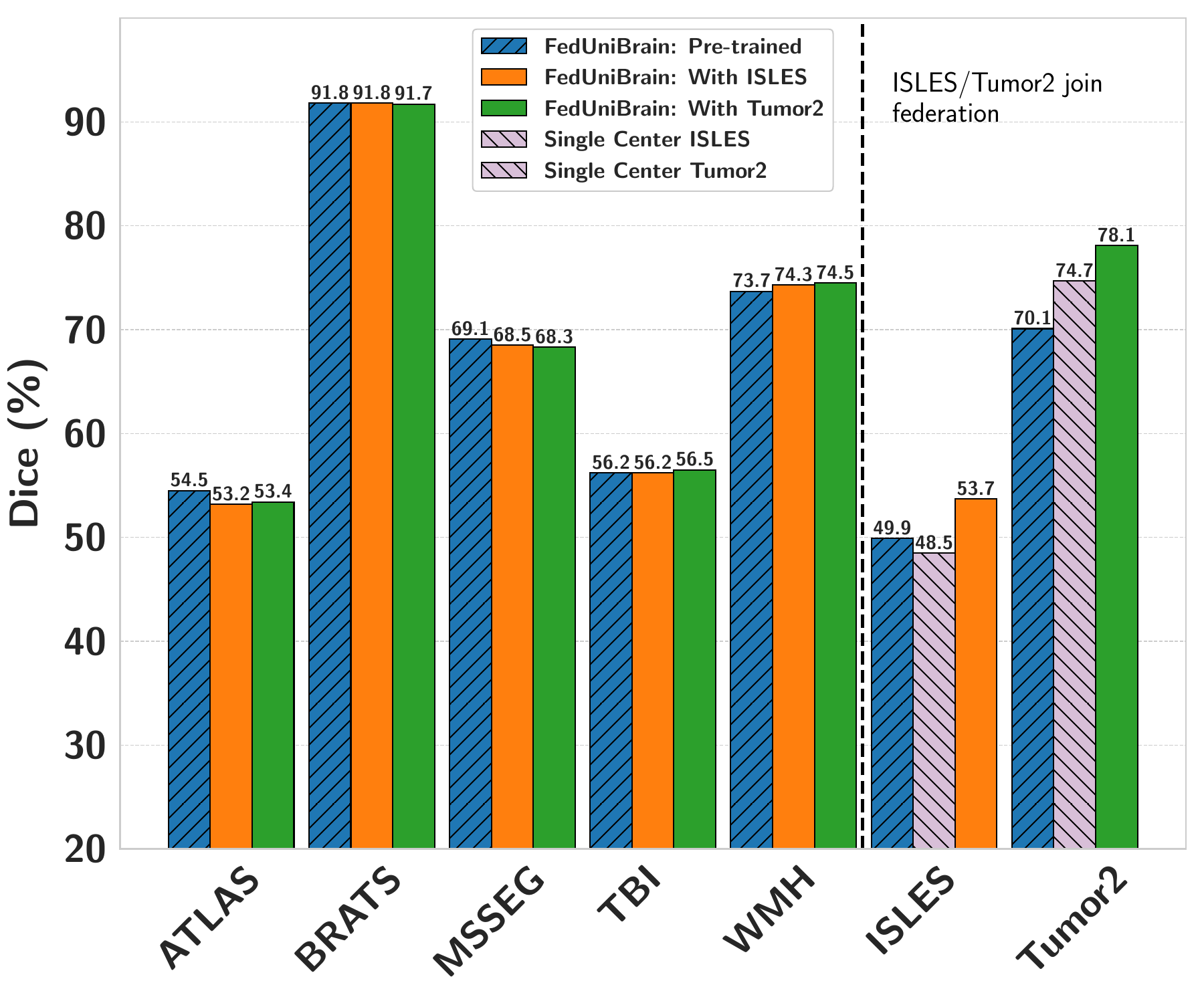}
        \caption{
        }
        \label{fig:ClientJoinsplot1}
    \end{subfigure}
    \begin{subfigure}[t]{0.31\textwidth}
        \centering
        \includegraphics[width=\linewidth]{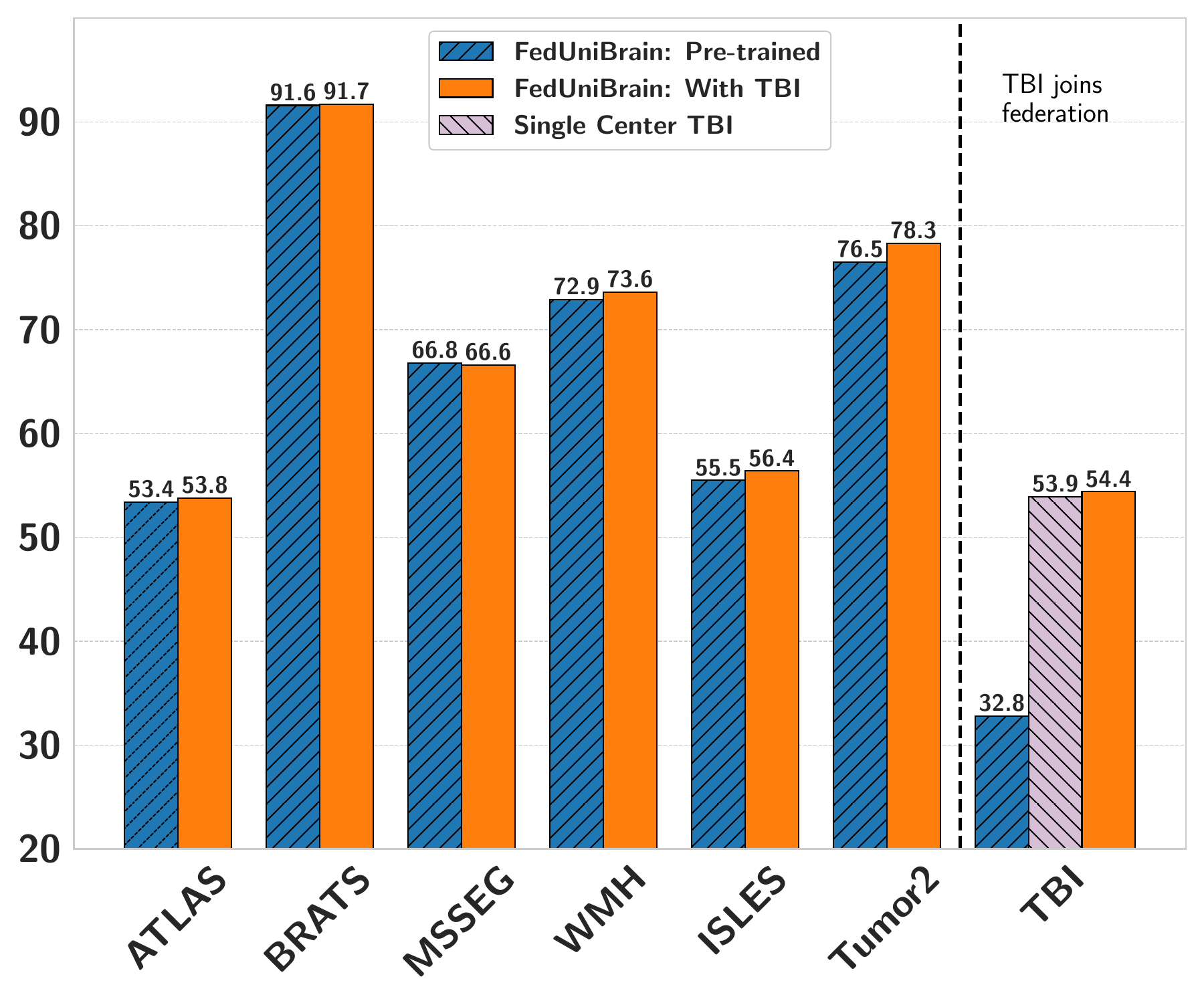}
        \caption{
        }
        \label{fig:ClientJoinsplot2}
    \end{subfigure}
    \begin{subfigure}[t]{0.31\textwidth}
        \centering
        \includegraphics[width=\linewidth]{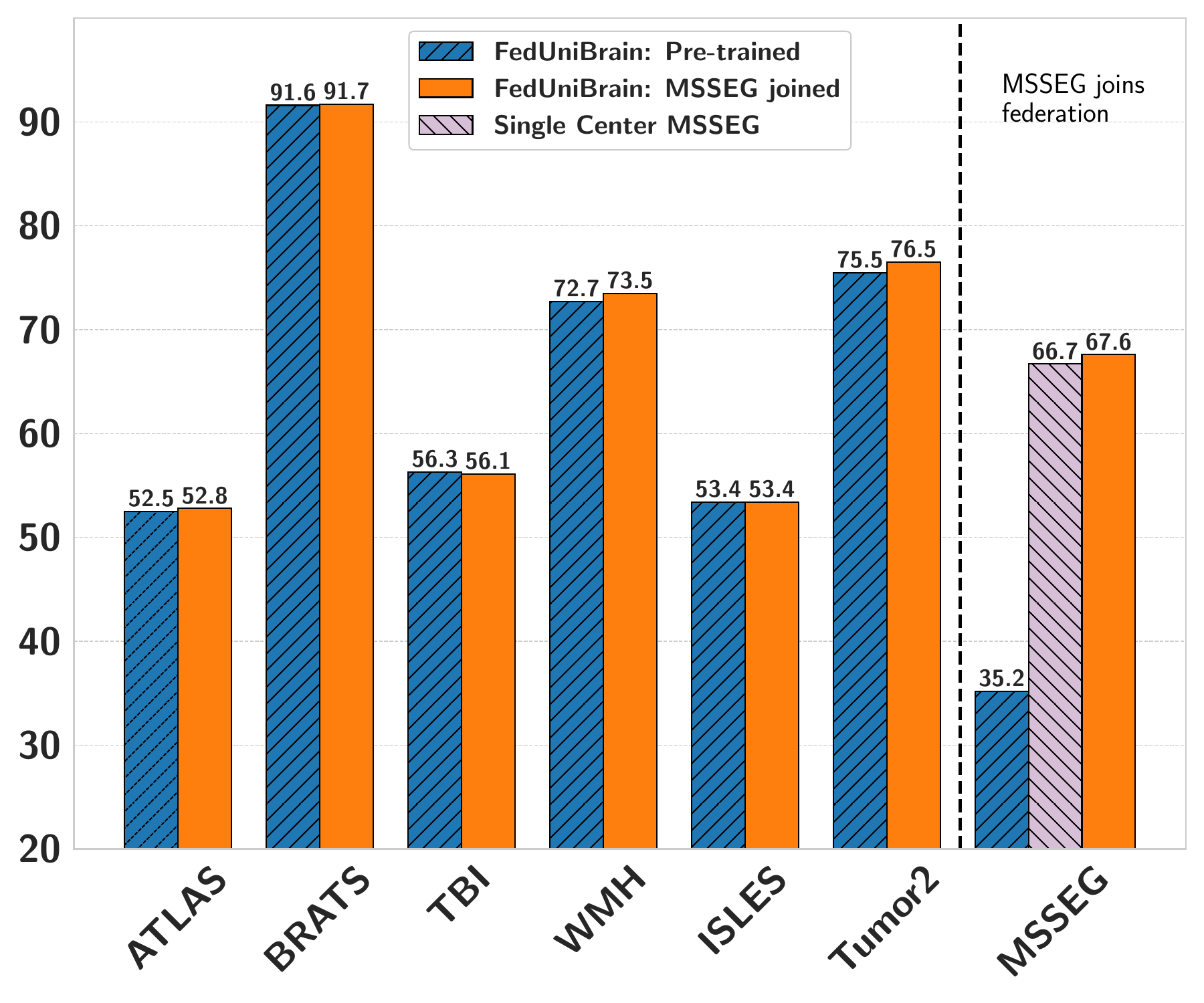}
        \caption{
        }
        \label{fig:ClientJoinsplot3}
    \end{subfigure}
    \caption{\textbf{Impact of a new client joining the federation.} Performance (Dice score \%) for both the initially trained clients and a client that joins at a later stage of training is presented. Blue bars represent the results before the new client joins (if the new client has an unseen modality, it is excluded during evaluation for the blue bar results). (a) Model initially trained on ATLAS, BRATS, MSSEG, TBI, and WMH (blue), then two scenarios are presented: after integrating ISLES client and learning its data (orange), and after integrating Tumor2 client (green). (b) Initial model trained without TBI (blue) and after integration and learning the TBI client (orange). (c) Initially trained without MSSEG (blue) and after integrating MSSEG (orange). Single-center training results on the new client are shown for comparison (violet).}
    \label{fig:ClientJoins}
\end{figure*}
\fi

\noindent \ul{\textbf{Effect of Feature Normalization on Generalization}}

\noindent Results in Tab.~\ref{tab:joint_training_results} indicate that learning client-specific BN parameters improves performance on source clients, compared to other normalization methods.
As shown in Tab.~\ref{tab:generalization}, however, when evaluating generalization on unseen databases (ISLES, Tumor2), BN with parameters tailored to source-clients underperforms compared to IN and NF.
Consequently, a trade-off must be made when designing a model for training with this type of heterogeneous data: either prioritize performance on source clients that contributed training data or generalize to unseen databases.

\noindent \ul{\textbf{Client with new modality and pathology joins federation}}

\noindent We evaluate the flexibility of our FedUniBrain framework in incorporating a new client that joins the federation at a later stage of training to further enhance the model with knowledge from new data. We consider three scenarios. \textbf{First}, a client with a \textbf{modality combination not seen during training} joins. Initially, the model is trained on ATLAS, BRATS, MSSEG, TBI, and WMH, then Tumor2 database joins (Fig.~\ref{fig:ClientJoinsplot1}). \textbf{Second}, a client with a \textbf{modality not seen during earlier training} joins, specifically the ISLES database that includes the unseen DWI modality (Fig.~\ref{fig:ClientJoinsplot1}). \textbf{Lastly}, a client with a \textbf{modality and pathology not seen during training} joins. We test this scenario in two cases: excluding TBI from initial training, the only database with SWI modality and Traumatic Brain Injuries  (Fig.~\ref{fig:ClientJoinsplot2}), and excluding MSSEG, which is the only database with PD modality and Multiple Sclerosis (Fig.~\ref{fig:ClientJoinsplot3}).

The FedUniBrain model is trained for 300 communication rounds before a new client joins, followed by 200 additional rounds (optimization as per~\ref{sub:Setup}).
Results in Fig.~\ref{fig:ClientJoins} show that the Dice scores for the 5 initial training databases remain stable after a new client joins and even slightly improve when a client with a previously unseen pathology joins (Fig.~\ref{fig:ClientJoinsplot2},~\ref{fig:ClientJoinsplot3}). Moreover, FedUniBrain learns to segment the new pathologies and modalities of new clients, outperforming training only on the specific client (single-center), as confirmed by our statistical superiority tests for pathologies seen during training ISLES (p=0.042) and Tumor2 (p=0.047), which indicates positive knowledge transfer. For the challenging setting of completely unseen modalities and pathologies of TBI (p$<$0.001) and MSSEG (p=0.006), non-inferiority tests show that FedUniBrain matches single-center performance. This demonstrates that our framework handles new clients seamlessly, without re-initialization or complex modifications, and not overfitting to previously learnt diseases.  
Experiments on zero-shot generalization (Tab.~\ref{tab:generalization}) show strong performance on unseen databases with pathologies seen during training. However, as shown by the blue bars in Fig.~\ref{fig:ClientJoinsplot2} and~\ref{fig:ClientJoinsplot3}, generalization struggles with pathologies not seen during training. 
The FedUniBrain framework can easily integrate new clients, even with previously unseen pathologies and modalities, and learn to effectively segment the new type of data.

\section{Conclusion}
This study demonstrated the feasibility and effectiveness of using Federated Learning to train a single segmentation model across decentralised brain MRI databases, each with different input modalities and pathologies. Our results show that federated joint training on multiple databases results in a \emph{single} model that can segment \emph{all} diseases and sets of modalities seen in training clients with performance similar to client-specific models tailored for a specific set of modalities and disease. Promising generalization to unseen clients with modality combinations not seen during training is also demonstrated. The study finally highlights the importance of choosing feature normalization method, revealing a possible trade-off between generalization to unseen databases or tailored performance for clients that contribute to training. 
Having demonstrated, for the first time, the feasibility of FL from clients with such diverse brain lesion MRI data, the study opens the possibility for real-world collaborative federations to be initiated for this application.

\section{Acknowledgement}

FW is supported by the EPSRC Centre for Doctoral Training in Health Data Science (EP/S02428X/1), by the Anglo-Austrian Society, and by an Oxford-Reuben scholarship.
PS is supported in part by the UK EPSRC Programme Grant EP/T028572/1 (VisualAI), in part by the UKRI grant reference EP/X040186/1 (Turing AI Fellowship: Ultra Sound Multi-Modal Video-based Human-Machine Collaboration), and a UK EPSRC Doctoral Training Partnership award.
ZL is supported by scholarship provided by the EPSRC Doctoral Training Partnerships programme [EP/W524311/1].
VN, NIHR Rosetrees Trust Advanced Fellowship, NIHR302544, is funded in partnership by the NIHR and Rosetrees Trust. 
NV is supported by the NIHR Oxford Health Biomedical Research Centre (NIHR203316). The views expressed are those of the authors and not necessarily those of the NIHR, Rosetrees Trust or the Department of Health and Social Care.
The authors also acknowledge the use of the University of Oxford Advanced Research Computing (ARC) facility in carrying out this work (http://dx.doi.org/10.5281/zenodo.22558). We would like to thank Abhirup Banerjee for his help with the statistical analysis in the Supplementary.

{\small
\bibliographystyle{ieee_fullname}
\bibliography{egbib}
}

\appendix
\renewcommand{\thesection}{\Alph{section}}
\renewcommand{\thefigure}{\thesection\arabic{figure}}
\renewcommand{\thetable}{\thesection.\arabic{table}}

\section{Algorithm}\label{sec:intro}
Alg.~\ref{alg:FedUniBrain} shows the pseudo-code of the FedUniBrain framework. We omit the mini-batch dimensionality for clarity.

\begin{algorithm}[h]
\small
\caption{FedUniBrain for Multi-modal MRI Seg.}
\label{alg:FedUniBrain}
\begin{algorithmic}[1]
\State \textbf{Input:} Set of clients $C$, each with dataset $\mathcal{D}_c$
\State \textbf{Output:} Trained global model parameters $\theta$
\Function{\textbf{Server executes:}}{}
    \State{$\triangleright$ \textit{First, get unique modalities and intialize}}
    \For{each client $c \in C$ \textbf{in parallel}}
        \State $\mathcal{M}_c \gets \Call{ClientGetModalities}{c}(\mathcal{D}_c)$
    \EndFor
\State $\mathcal{M} \gets \bigcup_{c \in C} \mathcal{M}_c$
\State Initialize global model $\theta$ with $|\mathcal{M}|$ input channels
    \For{$e=1$ to $E$} \Comment{\textit{Begin training}}
        \For{each client $c \in C$ \textbf{in parallel}}
            \State $\theta_c \gets \Call{ClientUpdate}{c,\theta}$
        \EndFor
        \For{each client model $c$ and each layer $l$}
            \If{keep client-specific BN params $= \text{True}$:}
                \If{$l \neq \text{BN layer}$:}
                    \State \(\theta^l = \frac{1}{C} \sum_{c=1}^{C} \theta^l_c\)
                \EndIf
            \Else :
                \State \(\theta^l = \frac{1}{C} \sum_{c=1}^{C} \theta^l_c\)
        \EndIf
        \EndFor
    \EndFor
\EndFunction
\LeftComment{Below is executed on the clients}
\Function{ClientGetModalities}{c}: \Comment{\textit{Run on client c}}
    \State $\mathcal{M}_c \gets$ Set of input MRI modalities from $\mathcal{D}_c$
    \State \textbf{return} $\mathcal{M}_c$ to server
\EndFunction
\Function{ClientUpdate}{$c,\theta, \mathcal{M}$}: \Comment{\textit{Run on client c}}
    \State $\theta_c \gets \theta$
    \If{client-specific BN params $= \text{True}$}:
        \State Overwrite all BN params with $\gamma_c, \beta_c, \hat{\mu}_c, \hat{\sigma}_c$
    \EndIf
    \For{$t = 1$ to $\tau$}
        \State \( d_c \sim \mathcal{D}_c \) \Comment{\textit{Sample mini-batch}}
        \State Initialize blank input tensor $b \in \mathbb{R}^{|\mathcal{M}|\times w \times h \times d}$
        \State{$\triangleright$ \textit{Copy client data to input tensor}}
        \State $b[i,:,:,:] \gets d_c[i,:,:,:] \quad \forall i \in \mathcal{M}_c$
        \State $k \gets \Call{randint}{1, |\mathcal{M}_c|}$
        \State $\mathcal{M}_b \gets$ Rand. sample $k$ modalities from $\mathcal{M}_c$
        \State $\triangleright$ \textit{Modality Drop by setting modalities blank}
        \State $b[j,:,:,:] \gets 0. \quad \forall j \in \mathcal{M}_b$
        \State Perform local update on $\theta_c$ using: $\mathcal{L}_c(\theta_c; b; \mathcal{T}_c)$
    \EndFor
    \State \textbf{return} $\theta_c$
\EndFunction

\end{algorithmic}
\end{algorithm}

\section{Additional Training Details}
\textbf{Additional Hyperparameter Group Norm (GN)}: In the results presented in Tab. 2, we use FedUniBrain with GN as an alternative non-client-specific feature normalization technique. The number of groups is an additional hyperparameter for GN, which we set to 16 for our experiments. 

\textbf{Centralised MultiUNet experiments:} For the MultiUNet results in Tab. 4 we use the same training setup as described in \cite{xu2024feasibility}. However, to enable a direct comparison, we use the same number of epochs, learning rates, and loss function as in our Federated Learning experiments.

\section{Datasets and their Different MRI Modalities}
\begin{figure}[h]
    \centering
    \includegraphics[width=\columnwidth]{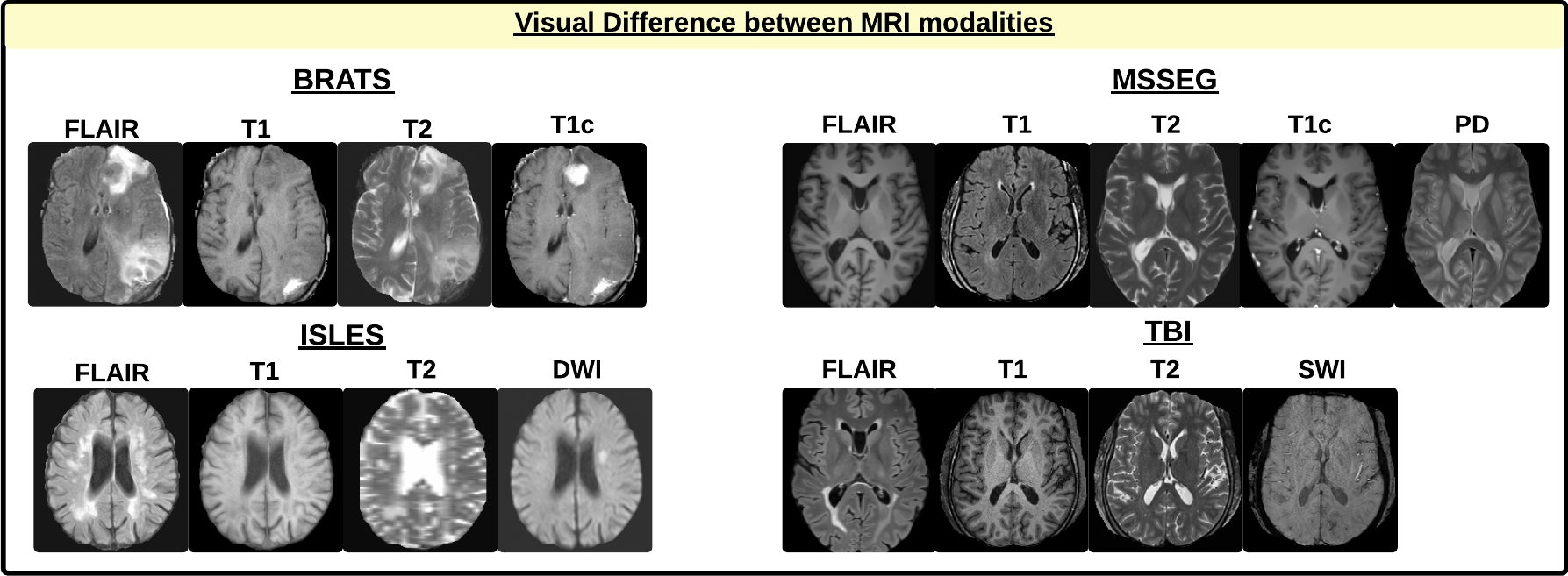}
    \caption{\textbf{MRI Databases with their different modalities.}}
    \label{fig:multiModal}
\end{figure}

In Fig.~\ref{fig:multiModal}, we show visual representations demonstrating that each MRI modality provides complementary information about tissues and anatomical structures through different contrasts. For each of the four databases - BRATS, ISLES, MSSEG, TBI - we show the same anatomical slices across different modalities, illustrating that each modality offers different diagnostic insights into the brain.

\section{Statistical Analysis}\label{sup:sec:stats}
The primary goal of this paper is to investigate the feasibility of training a single federated model, FedUniBrain, that can effectively segment multiple diseases across different databases, each with different MRI input modalities and brain pathologies. Since this approach has not been previously explored for federated training, showing that FedUniBrain does not overfit to a single dataset, disease, or input modality combination would be a great success. This would show the feasibility and benefits of training a unified model.

In some instances, our goal is to show that FedUniBrain performs at least as well as a centralised model, through non-inferiority testing, which would confirm the feasibility of training a single federated model. In other cases, we aim to show that FedUniBrain outperforms single-center training, through superiority tests, highlighting additional benefits of federated training a unified model.

Since we are working with segmentation models, each method generates a distribution of Dice scores per experiment (one Dice score per patient). This allows us to compute the Dice scores of the test data in each experiment for each method and conduct one-sided t-test for non-inferiority or superiority, depending on our claim. Following, we explain the non-inferiority test and superiority test used in this analysis.

\noindent \textbf{\underline{Non-inferiority test:}} 
To evaluate whether the federated approach, FedUniBrain, is not inferior to a specific method (based on our claim), we conduct statistical non-inferiority tests. In these tests, the null hypothesis ($H_0$) states that the mean difference in Dice performance between FedUniBrain and the specific method is less than or equal to a specified margin, indicating the FedUniBrain is inferior. The alternative hypothesis (\(H_1\)) states that the mean difference in Dice performance is greater than \(-\Delta\), suggesting that FedUniBrain is not worse than the specific method by more than the margin \(\Delta\). 
\begin{equation*}
\begin{aligned}
H_0 &: \delta_d \leq - \Delta \\
H_1 &: \delta_d > - \Delta,
\end{aligned}
\end{equation*}
where $\delta_d$ represents the mean paired differences of the samples' Dice scores, $\Delta$ represents a pre-specified margin, which we set to $5\%$, which is a commonly used value \cite{mckinney2020international}. To conduct the non-inferiority tests we, use a \textbf{paired one-sided t-test}.

\noindent \textbf{\underline{Superiority test:}}
To evaluate if the federated approach FedUniBrain is better than specific method (based on our claim) in terms of segmentation performance, we employ statistical superiority tests. We also use a \textbf{paired one-sided t-test} to evaluate whether the mean performance of FedUniBrain is superior to the mean performance of a specific method. The null hypothesis ($H0$) states that the mean difference in Dice performance is less than or equal to zero, indicating that FedUniBrain is not superior. The alternative hypothesis ($H1$) states that the mean difference in performance is greater than zero, showing that FedUniBrain's performance is superior compared to the specific method:
\begin{equation*}
\begin{aligned}
H_0 &: \delta_d \leq 0 \\
H_1 &: \delta_d > 0.
\end{aligned}
\end{equation*}

\noindent \textbf{\underline{Reporting of the statistical results:} }For all our tests, we report the number of samples in our test (N), the difference in means, the 95\% confidence intervals, and the p-value.

Note, we are reporting the 95\% confidence intervals for the difference in means of the Dice performance (from the paired t-test). For successful \textbf{superiority} testing, positive confidence intervals are expected, and a confidence interval's lower limit above 0 demonstrates statistical superiority. If the 95\% confidence interval lower limit is below zero, it shows no statistically superiority. However, for non-inferiority testing, the interpretation is different. Here, we want to show that the mean performance difference is not worse than our predefined margin $\Delta$. This means that negative confidence intervals are acceptable and indicate statistical significance, as long as the lower bound does not fall below the margin $-\Delta$.
Note, since we use a one-sided test, it is expected that the 95\% confidence interval has an upper bound of infinity. 

We reject the null hypothesis if the p-value falls below 0.05 (our significance level), which would indicate that FedUniBrain is statistically not-inferior (= not worse than) or superior (= better than), depending on the test.

\begin{table*}[!h]
\centering
\caption{Statistical comparison of FedUniBrain with single center training for segmentation performance}
\label{tab:comparisonFedVsSingleCenter}
\resizebox{\textwidth}{!}{
\begin{tabular}{lccccccc}
\toprule
\textbf{Client joining} & \textbf{Stat. Test} & \textbf{Samples (N)} & \textbf{Mean Federated Dice} & \textbf{Mean Single-Center Dice} & \textbf{Mean of Diff.} & \textbf{95\% CI} & \textbf{p-Value} \\
\midrule
ATLAS & Superiority & 195 & 54.5 & 52.8 & 1.73 & (0.27, $\infty$) & \textbf{0.0260} \\
BRATS & Non-Inferiority & 40 & 91.8 & 91.9 &  -0.13 & (-0.71, $\infty$) & \textbf{2.2e-16} \\
MSSEG & Non-Inferiority & 16 & 69.1 & 68.3 &  0.79 & (-3.16, $\infty$) & \textbf{0.0106} \\
TBI & Non-Inferiority & 125 & 56.2 & 56.1 &  0.11 & (-1.03, $\infty$) & \textbf{8.538e-12} \\
WMH & Superiority & 18 & 73.7 & 71.5 & 2.21 & (1.19, $\infty$) & \textbf{7.335e-06} \\
\bottomrule
\end{tabular}
}
\end{table*}

\begin{table*}[!h]
\centering
\caption{Statistical comparison of FedUniBrain with the centralized method for segmentation performance on ISLES and Tumor2}
\label{tab:comparisonFedVsCent}
\resizebox{\textwidth}{!}{
\begin{tabular}{lccccccc|ccccccc} 
\toprule
& \multicolumn{6}{c}{\textbf{ISLES}} & \multicolumn{7}{c}{\textbf{Tumor2}} \\
\cmidrule(lr){3-8} \cmidrule(lr){9-14}
\textbf{Model} & \textbf{Norm} & \textbf{N} & \textbf{Cent. Mean Dice} & \textbf{Fed. Mean Dice} & \textbf{Mean of Diff.} & \textbf{95\% CI} & \textbf{p-Value} & \textbf{N} & \textbf{Cent. Mean Dice} & \textbf{Fed. Mean Dice} & \textbf{Mean of Diff.} & \textbf{95\% CI} & \textbf{p-Value} \\
\midrule
\underline{FedUniBrain} & IN & 28 & 55.5 & 55.3 & 0.26 & (-2.71, $\infty$) & \textbf{0.0027} & 57 & 72.2 & 72.7 & 0.12 & (-1.45, $\infty$) & \textbf{5.443e-07} \\
FedUniBrain & NF & 28 & 55.5 & 52.8 & -2.18 & (-6.01, $\infty$) & 0.1105 & 57 & 72.2 & 72.1 & -0.53 & (-1.65, $\infty$) & \textbf{6.181e-09} \\
FedUniBrain (avg. BN params) & BN & 28 & 55.5 & 54.5 & -0.49 & (-4.91, $\infty$) & \textbf{0.0469} & 57 & 72.2 & 68.0 & -4.62 & (-6.76, $\infty$) & 0.3848 \\
FedUniBrain (client spec. BN params) & BN & 28 & 55.5 & 49.9 & -5.1 & (-11.31, $\infty$) & 0.5079 & 57 & 72.2 & 70.1 & -2.52 & (-3.91, $\infty$) & \textbf{0.0021} \\
\bottomrule
\end{tabular}
}
\end{table*}

\begin{table*}[!h]
\centering
\caption{Statistical comparison of FedUniBrain models with single center training when a new client joins in for segmentation performance}
\label{tab:comparisonNewClient}
\resizebox{\textwidth}{!}{
\begin{tabular}{lccccccc}
\toprule
\textbf{Client joining} & \textbf{Stat. Test} & \textbf{Samples (N)} & \textbf{Mean Single-Center Dice} & \textbf{Mean Federated Dice} & \textbf{Mean of Diff.} & \textbf{95\% CI} & \textbf{p-Value} \\
\midrule
ISLES & Superiority & 8 & 48.5 & 53.7 & 4.79 & (0.24, $\infty$) & \textbf{0.0422} \\
Tumor2 & Superiority & 17 & 74.7 & 78.1 &  3.15 & (0.07, $\infty$) & \textbf{0.0466} \\
TBI & Non-Inferiority &125 & 53.9 & 54.4 &  0.51 & (-1.06, $\infty$) & \textbf{2.399e-08} \\
MSSEG & Non-Inferiority & 16 & 66.7 & 67.6 &  0.91 & (-2.72, $\infty$) & \textbf{0.0061} \\
\bottomrule
\end{tabular}
}
\end{table*}

\subsection{Statistical Analysis of FedUniBrain and Single Center}

In this section, we statistically test FedUniBrain against single-center training. Our goal is to demonstrate that FedUniBrain can match or even improve the performance of single-center training, which would be a major success as it would show that our model does not overfit to a specific dataset, disease, or input modality combination. Achieving this would validate the main goal of our paper, proving that training a single model across multiple databases with different diseases and input modality combinations is feasible. We show that FedUniBrain is either statistically superior or non-inferior compared to single-center training for all datasets.

The results are presented in Table \ref{tab:comparisonFedVsSingleCenter}, corresponding to the results of Tab. 2 of the main paper. We compare FedUniBrain with client-specific batch normalization (BN) parameters and modality drop against single-center training \textbf{without} modality drop (because these two are the best performing approaches). The results indicate that FedUniBrain consistently performs non-inferior than single-center training and, for the ATLAS and WMH datasets superior.

\subsection{Statistical Analysis of Zero-Shot Generalization}

This section shows a statistical analysis comparing FedUniBrain with different normalization techniques to the centralized MultiUnet approach (all of them \textbf{with} Modality Drop). Demonstrating that FedUniBrain is statistically not-inferior than the centralized method, MultiUNet, would be a major accomplishment, indicating that FedUniBrain is on par with the centralized method. We perform non-inferiority tests between FedUniBrain with different normalization methods and the centralized MultiUNet method. These tests correspond to the results from Tab. 4 of the main paper.

The results in Tab. \ref{tab:comparisonFedVsCent} show that FedUniBrain with Instance Normalization (IN) is significantly not-inferior to the centralized method for both the ISLES and Tumor2 datasets. Comparing different normalization techniques for FedUniBrain, IN is the only one that is statistically non-inferior on both datasets and therefore comes closest to matching the centralized method. This strengthens our argument that the choice of normalization is important depending on whether the goal is a personalized model or a model that generalizes well.

\subsection{Statistical Analysis of a new client joining the federation}

In this section, we want to evaluate whether FedUniBrain does not perform worse to or even surpass single-center training in the challenging scenario where a new client joins during training. Specifically, for scenarios, where a new client with a known pathology joins, we perform statistical superiority tests. The results of these tests are presented in Tab. \ref{tab:comparisonNewClient} and correspond to the results shown in Fig. 4 of the main paper. From the results, we can see that when a new client with an already seen pathology joins, FedUniBrain performs significantly better than single-center training (Tumor2 and ISLES).
In the more challenging scenario where a new client with a previously unseen pathology, \textbf{and} unseen modality joins, we perform statistical non-inferiority tests. The results of these tests demonstrate that FedUniBrain is statistically not inferior to single-center training, which is an important results. This indicates that even in the highly challenging setting of continual learning, FedUniBrain does not overfit to specific databases and is capable of learning from a completely new database with a new brain pathology including a new modality, as effectively as single-center training.

\end{document}